\documentclass[useAMS,usenatbib]{mn2e}
\usepackage{epsfig}
\title[Anisotropy of E-Galaxies]
{The Surprising Anisotropy of Fast Rotating, Disky Elliptical Galaxies}
\author[Andreas Burkert \& Thorsten Naab]
  {Andreas ~Burkert and
   Thorsten ~Naab\\
University Observatory Munich, Scheinerstr. 1, D-81679 Munich, Germany \\
   }
\date{Submitted to MNRAS}

\pagerange{\pageref{firstpage}--\pageref{lastpage}} \pubyear{2005}

\def\LaTeX{L\kern-.36em\raise.3ex\hbox{a}\kern-.15em
    T\kern-.1667em\lower.7ex\hbox{E}\kern-.125emX}

\begin{document}

\label{firstpage}

\maketitle

\begin{abstract}

The projected kinematical properties of unequal-mass merger remnants of disk galaxies
are analysed and shown to agree well with observations of disky, fast rotating elliptical galaxies. 
This supports the major merger hypothesis of early-type galaxy formation. However,
in contrast to previous claims, the merger remnants are very anisotropic 
with values of the anisotropy parameter that are similar to equal-mass merger 
remnants that form boxy, slowly rotating ellipticals. Including gas in the 
simulations does not change this result although the line-of-sight
velocity profile and the intrinsic orbital structure are strongly affected by the presence 
of gas. 
The kinematical difference between boxy and disky ellipticals appears not to be the
amount of anisotropy but rather rotation and the shape of the velocity dispersion 
tensor. The apparent isotropy of observed disky ellipticals is shown to result from
inclination effects. Even small inclination angles strongly reduce
the measured anisotropy of fast rotating systems, seen in projection.
A second problem is the limited amount of information that is available 
when measuring only the central velocity dispersion and a characteristic rotation and ellipticity.
Methods are investigated that allow a better determination of the intrinsic
anisotropy of fast rotating early-type galaxies with known inclination angles.
\end{abstract}


\section{Introduction}

Elliptical galaxies are old stellar systems that are believed to have
formed from major mergers of disk galaxies, preferentially
at high redshifts (Searle, Sargent \& Bagnuolo 1973; Toomre \& Toomre 1972).
This ``merger hypothesis``  has been tested using numerical simulations
(e.g. Gerhard 1981, Negroponte \& White 1983, Barnes 1988). Early simulations
demonstrated consistently that merger remnants can account for
many characteristic global properties of ellipticals, like 
their surface density profiles (Trujillo et al. 2004), ellipticities,
sizes and large velocity dispersions.

More recently it has however become clear that early-type galaxies are more complex than 
originally thought.  Isophotal fine structures have been detected that correlate well with 
kinematical properties (Bender et al. 1989).  Faint ellipticals are fast rotators with 
small minor axis rotation. They are called disky as a Fourier analyses of their isophotal 
deviations from perfect ellipses leads to positive values of the fourth order Fourier coefficient 
$a_4$. In contrast, bright ellipticals are in general boxy with negative values of 
$a_4$ and slow rotation. The importance of anisotropy has been estimated using the
so called anisotropy parameter (Binney 1978)

\begin{equation}
\left( \frac{v}{\sigma} \right)^* = \frac{v/\sigma}{\sqrt{\epsilon/(1-\epsilon)}},
\end{equation}

\noindent where $v$ is a characteristic rotational velocity, $\sigma$ is the central velocity dispersion,
and $\epsilon$ is the ellipticity.  Disky ellipticals have values of $(v/\sigma)^* \geq 0.7$ and therefore
are believed to be rotationally flattened, isotropic stellar systems. Boxy ellipticals,
on the other hand, are characterized by $(v/\sigma)^* << 1$, indicating that they are
flattened by an anisotropic velocity dispersion.

In order to understand the origin of boxy and disky ellipticals and their kinematical
properties within the framework of the major merger scenario many simulations have
been performed (Hernquist 1992, 1993; Naab, Burkert \& Hernquist 1999; 
Bendo \& Barnes 2000; Gonz\'{a}lez-Garc\'{i}a \& van Albada 2003; Nipotti, Londrillo \& Ciotti 2003; 
for recent reviews see Burkert \& Naab 2004a,b). 
Naab \& Burkert (2003) presented a large parameter survey of disk galaxy
mergers with statistically unbiased orbital initial conditions and different 
mass ratios $\eta$ of the progenitors.
They showed that unequal mass 3:1 and 4:1 mergers lead to fast rotating,
disky systems, in good agreement with the observations of disky ellipticals.
Equal-mass mergers, on the other hand, tend to form slowly rotating, boxy ellipticals.
2:1 mergers generate a mixed population of boxy and disky objects.
Most massive ellipticals are boxy, while
2/3 of the lower-mass ellipticals are disky (Bender, Burstein \& Faber 1992). This observation
is at first not expected within the framework of a scenario where the isophotal
shape depends mainly on the mass ratio $\eta$ of the merging galaxies
as cosmological models predict that the distribution of $\eta$ is independent of galaxy 
mass (Khochfar \& Burkert 2005). Khochfar \& Burkert (2005) however showed
that the observations can be reconciled with theory if mixed mergers between
ellipticals and spirals and elliptical-elliptical mergers are taken
into account (see also Gonz\'{a}lez-Garc\'{i}a \& Balcells 2005) which dominate for high
masses and produce boxy, anisotropic remnants, independent of $\eta$.

In summary,  a consistent picture of early-type galaxy formation is 
emerging and theoretical investigations are now focussing on a more 
detailed understanding of star formation
and energetic feedback processes during major mergers (Bekki 1999, Mihos \& Hernquist 1996,
Meza et al. 2003; Cox et al. 2005). A particularly
interesting recent study in this context is the growth of central black holes or
AGN feedback and its effect on the gaseous and stellar environment
(Merrifield 2004; Springel, Di Matteo \& Hernquist 2005; Di Matteo, Springel \& Hernquist 2005)
Although feedback is treated very simplified, these models can
successfully reproduce the tight correlation between the mass of central black holes
and the velocity dispersion of the stellar component (Ferrarese \& Merritt 2000; Gebhardt et al. 2000;
Tremaine et al. 2002).

Despite all of this progress a still unsolved puzzle is the apparent isotropy of fast rotating, 
disky ellipticals.  Major mergers disturb kinematically cold, rotationally supported disk galaxies enough to
generate kinematically hot spheroidal stellar components. They should also
destroy any initially isotropic velocity distribution, leading at the end to anisotropic
systems. Two-body relaxation which drives the systems towards an isotropic velocity distribution 
is not efficient 
in ellipticals because of their long relaxation timescales that by far exceed the age of the Universe. 
In addition, Dehnen \& Gerhard (1994) showed that the line-of-sight velocity dispersions of fast
rotating ellipticals are not consistent with isotropic rotator models.
Why then do observed ellipticals appear isotropic and how does their hidden
anisotropy differ from that of boxy, slowly rotating ellipticals?

It is this question which we plan to investigate in this paper.
Section 2 summarizes some important stellar dynamical relations and definitions, following
the recent work of Binney (2005).  Section 3 investigates the distribution of observed ellipticals
in the anisotropy diagram. Section 4  demonstrates that unequal-mass mergers with and 
without gas lead to fast rotating ellipticals that are however very anisotropic in apparent
conflict with the observations. 
In section 5 inclination effects are investigated which can 
reconcile the theoretical models with the observations. 
Section 6 investigates the intrinsic anisotropy of merger remnants and compares
it with their apparent anisotropy, derived from edge-on projections.
A discussion of the results and conclusions follow in section 7.

\section{Analytical considerations}

It is a fundamental stellar dynamical problem to derive the intrinsic dynamical state
of a stellar system from its observed projected properties. Binney (1978)
used the tensor virial theorem to demonstrate that observed
massive, slowly rotating elliptical galaxies are shaped by an
anisotropic velocity distribution. Recently, 
Binney (2005) presented a revised version of his earlier work which properly
takes into account projection effects.
Here, we shortly summarize the calculations of Binney (2005) which will provide 
the theoretical basis for the subsequent analysis of the merger remnants. We will
restrict ourselves to oblate systems which, as shown below, describe well the geometry
of disky, unequal mass merger remnants.

Consider an axisymmetric, oblate stellar system with density distribution $\rho(\vec{x})$. The
lengths of the semi-axes of its equidensity surfaces are denoted as $a_i$. If the x-y-plane is the 
equatorial plane, $a_x=a_y$ and $a_z < a_x$. We assume that the dominant mean stellar 
motion is rotation around the z-axis. The equilibrium state of the system is 
completely determined by the stellar phase space distribution function 
$f(\vec{x},\vec{v})$ d$^3$x d$^3$v which denotes the number of stars at location 
$\vec{x}$ with velocity $\vec{v}$ in the phase space volume d$^3$x d$^3$v.
Given $f$, the density of stars at $\vec{x}$ is given by

\begin{equation}
\rho (\vec{x}) = \int f \ \mathbf{d^3v}
\end{equation}

\noindent and the mean stellar streaming velocity in the i-th direction is 

\begin{equation}
\bar{v}_i(\vec{x}) = \frac{1}{\rho} \int v_i f \ \mathbf{d^3v} .
\end{equation}

\noindent The local velocity dispersion in the i-th direction is

\begin{equation}
\sigma_i^2 (\vec{x}) = \frac{1}{\rho} \int (v_i - \bar{v}_i)^2 f \ \mathbf{d^3v}.
\end{equation}

\noindent For oblate systems $\sigma_x = \sigma_y \neq \sigma_z$.
We now can define the anisotropy $\delta$ of the system 

\begin{equation}
\delta \equiv \frac{\Pi_{xx} - \Pi_{zz}}{\Pi_{xx}}
\end{equation}

\noindent where the

\begin{equation}
\Pi_{ii} \equiv \int \rho \sigma^2_i \mathbf{d^3x}
\end{equation}

\noindent are the diagonal elements of the velocity dispersion tensor $\Pi_{ij}$. In the
isotropic case $\delta = 0$ and $\Pi_{xx} = \Pi_{yy} = \Pi_{zz}$.

Stellar systems are observed in projection. The most ideal situation is the case 
of an edge-on system where the line-of-sight is parallel to the equatorial plane
(x-y plane). In the following we will assume that the y-axis
is in the direction of the line-of-sight.
Spectroscopic observations then allow a measurement of the
mass weighted projected velocity field 

\begin{equation}
\bar{v}_{los}(x,z) = \frac{1}{\Sigma} \int \int v_y f \ \mathbf{d^3v} \ \mathbf{dy}
\end{equation}

\noindent and the mass weighted projected velocity dispersion

\begin{equation}
\sigma^2_{los}(x,z) = \overline{v^2_{los}}(x,z) - \bar{v}^2_{los}(x,z) .
\end{equation}

\noindent where 

\begin{equation}
\overline{v^2_{los}}(x,z) = \frac{1}{\Sigma} \int \int v^2_y f \ \mathbf{d^3v} \ \mathbf{dy} 
\end{equation}

\noindent is the mean-square line-of-sight velocity and 

\begin{equation}
\Sigma (x,z) = \int \rho \ \mathbf{dy}
\end{equation}

\noindent is the surface density at $(x,z)$. 

Two characteristic, global kinematical observables can now be defined that
describe the kinematical state of the system: the mean-squared ordered motion

\begin{equation}
\langle \bar{v}^2_{los} \rangle = \frac{1}{M} \int \int \Sigma \ \bar{v}^2_{los} \ \mathbf{dx \ dz}
\end{equation}

\noindent and the mean squared velocity dispersion

\begin{equation}
\langle \sigma^2_{los} \rangle = \frac{1}{M} \int \int \Sigma \ \sigma^2_{los} \ \mathbf{dx \ dz}.
\end{equation}

\noindent where

\begin{equation}
M = \int \int \Sigma \ \mathbf{dx \ dz}
\end{equation}

\noindent is the total mass of the system.

Binney (2005) uses the virial theorem to derive a relationship between these 
parameters and the anisotropy $\delta$:

\begin{equation}
\left(\frac{v}{\sigma}\right)_{2d}^2 \equiv \frac{\langle \bar{v}^2_{los} \rangle}{\langle \sigma^2_{los} \rangle} = 
\frac{(1 - \delta ) W_{xx}/W_{zz} -1}{\alpha (1-\delta)W_{xx}/W_{zz}+1} .
\end{equation}

\noindent where

\begin{equation}
\alpha = \frac{1}{M \langle \bar{v}^2_{los} \rangle} \int u^2 \rho \ \mathbf{d^3x}
\end{equation}

\noindent measures the shear in the stellar streaming velocity and

\begin{equation}
u(\vec{x})=\bar{v}_y(\vec{x})-v_{los}(x,z) 
\end{equation}

\noindent is the difference between the streaming
velocity parallel to the line-of-sight at position $\vec{x}$ and the mean projected
line-of-sight velocity $v_{los}$ at $(x,z)$.
$W_{ii}$ is a diagonal element of the potential energy tensor.
When a system's equidensity surfaces are similar ellipsoids, $W_{xx}/W_{zz}$ depends only
on its eccentricity $\epsilon = (1-a_z/a_x)$, independent of the density distribution
$\rho$ (Roberts 1962, Binney 1978)

\begin{equation}
\frac{W_{xx}}{W_{zz}} \equiv q(e) = \frac{0.5}{1-e^2} \times \frac{\arcsin e - e \sqrt{1-e^2}}
{\frac{e}{\sqrt{1-e^2}}- \arcsin e}
\end{equation}

\noindent with

\begin{equation}
e = (1-(1- \epsilon )^2)^{1/2} .
\end{equation}

\begin{figure}
\vspace{1cm}
\centerline{\includegraphics[angle=0,scale=0.37]{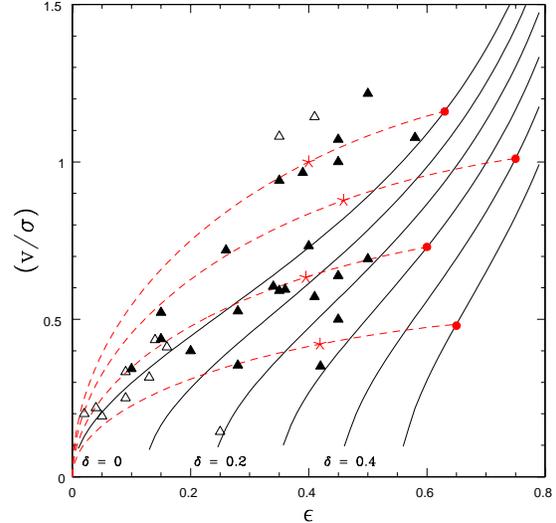}}
\caption{The anisotropy diagram. Solid lines show $(v/\sigma)_{2d}$ versus $\epsilon$ for
a given anisotropy $\delta$, adopting $\alpha = 0$. Red dashed curves show the effect of inclination
for systems which for edge-on projections ($\vartheta = 0$) are located at the red filled points
at the right end of each curve. Red stars indicate the locations for an inclination angle
of $\vartheta = 30^{\circ}$. Filled and open triangles show $(v/\sigma)_{1d}$ versus $\epsilon$
of observed disky ellipticals with $a_4 \geq 0.5$ and $0 \leq a_4 \leq 0.5$, respectively.}
\end{figure}

\noindent The solid lines in Fig. 1 (which we will call the anisotropy diagram)
show the relationship between $(v/\sigma)_{2d}$ 
and $\epsilon$ for different values of $\delta$, adopting $\alpha = 0$.
The larger $(v/\sigma)$ and the larger $\delta$, the larger the ellipticity. 

In principle, the equations (11) - (18) can be used to determine the anisotropy of oblate stellar systems
from 2-dimensional spectroscopic observations (Emsellem et al. 2004, de Zeeuw et al. 2002). 
In reality however, inclination effects need to be taken
into account (Binney \& Tremaine 1987). If $\vartheta$ denotes the
angle between the system's equatorial plane and the line of sight, the inclined
values are

\begin{eqnarray}
\epsilon_{inc} & = & 1-\sqrt{1-\epsilon (2- \epsilon) \cos^2 \vartheta} \\
\left(\frac{v}{\sigma}\right)_{inc} & = & \left(\frac{v}{\sigma}\right) 
\times \frac{\cos \vartheta}{(1- \delta \sin^2 \vartheta)^{1/2}} .
\end{eqnarray}

\noindent The dashed lines in Fig. 1 show how inclination affects $(v/\sigma)$ and
$\epsilon$. For large $(v/\sigma) > 0.5$ or large $\delta$ inclination decreases
 the observed ellipticity
with small changes in $(v/\sigma)$ as the inclination
curves cross the lines of constant $\delta$ at a large angle. Even small
inclination angles therefore can lead to a significant underestimate of the intrinsic anisotropy.
This will be crucial when we compare observations with numerical merger remnants.
As an illustration, the stars on each dashed curve show the location for an
inclination angle of $\vartheta = 30 ^{\circ}$, which is expected on average.

\section{Comparison with observed ellipticals}

Up to now ellipticals have mainly been observed along their apparent major and minor axis which does not 
allow us to derive
$\langle \bar{v}^2_{los} \rangle$ or $\langle \sigma^2_{los} \rangle$, required
to determine $\delta$ from equation (14). Instead the central line-of-sight velocity dispersion $\sigma_0$
and the characteristic peak rotational velocity along the major axis $v_{maj}$ have been used. In this case,
$\delta$ can be derived from the relation (Binney 1978, 2005)

\begin{equation}
\left( \frac{v}{\sigma} \right)_{1d}^2 \equiv \left( \frac{v_{maj}^2}{\sigma_0^2} \right) =
\frac{\pi^2}{8} \left( (1-\delta)\frac{W_{xx}}{W_{zz}}-1 \right) .
\end{equation}

Equation 21 is very similar to equation (14) for the case of $\alpha=0$ with the observational 
errors being much larger than the correction factor $(\pi^2 /8)$. We therefore 
will use Fig. 1 to investigate the anisotropy of observed ellipticals.

The filled triangles in Fig. 1 show observed disky elliptical galaxies 
with $a_4 \geq 0.5$ (Bender, Burstein \& Faber 1992).
Open triangles represent ellipticals with small values of $0 < a_4 < 0.5$ which populate the
lower left corner where the ellipticities and the rotational velocities are small.
This would be expected if these galaxies are seen almost face-on ($\vartheta = 90 ^{\circ})$ as
large inclinations lead to small $\epsilon$ and $(v/\sigma)$ and also significantly
reduce the projected $a_4$-values.
At first, the observations appear to support the standard expectation
that disky ellipticals are on average isotropic ($\delta = 0$) and rotationally flattened.

\section{The location of merger remnants in the anisotropy diagram}

Naab \& Burkert (2001, 2003) presented a large parameter set of collisionless galaxy mergers.
Equilibrium spirals were generated following Hernquist (1993), 
consisting of an exponential disk, a spherical, non-rotating
bulge and a pseudo-isothermal halo. The mass ratios $\eta$ of the progenitor disks were varied
between $\eta = 1$ and $\eta = 4$. The galaxies were assumed to approach each other on nearly
parabolic orbits with an initial separation of 30 length units and a pericenter distance of
2 length units, where a length unit is equal to the exponential scale length of the more massive
disk galaxy. Free parameters were the inclinations of the two disks relative to the orbital
plane and the arguments of pericenter. In order to select an unbiased sample of initial
disk orientations the procedure described by Barnes (1988) was applied. In total 16
equal-mass mergers and 96 mergers with $\eta$ = 2,3,4 were calculated. The merger products were
allowed to settle into equilibrium for 10 dynamical timescales. 

The remnants were analysed following
as closely as possible the procedures of observers as described by Bender (1988a,b).
First, an artificial image of the remnant was created by binning the projected remnant 
into $128 \times 128$ pixels. This picture was smoothed with a Gaussian filter of standard deviation 
1.5 pixels. The isophotes and their $a_4$ values
were then determined using a data reduction package provided by Ralf Bender. Following
the standard definitions of observers, the central projected velocity dispersion 
$\sigma_0$ inside a projected galactocentric distance of 
0.2 effective radii, the characteristic ellipticity $\epsilon$, defined as the isophotal
ellipticity at 1.5 effective radii and the characteristic rotational velocity $v_{maj}$,
defined as the projected rotational velocity on the major axis at 1.5 effective radii was
determined.

\begin{figure}
\vspace{1cm}
\centerline{\includegraphics[angle=0,scale=0.37]{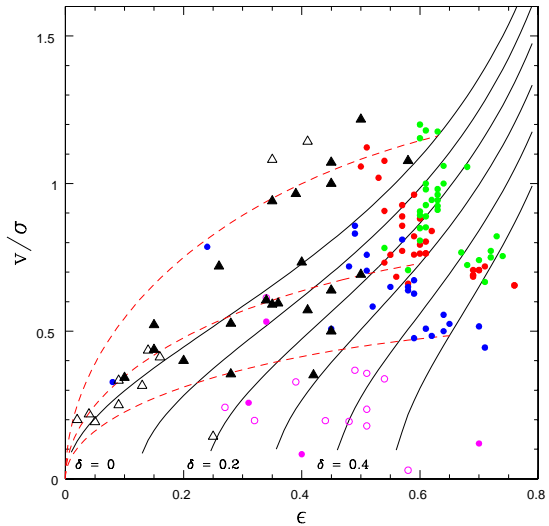}}
\caption{$/v/\sigma )_{1d}$ versus $\epsilon$ 
of edge-on ($\vartheta = 0$) merger remnants in the anisotropy diagram.
Green points show 4:1 mergers, red points correspond to 3:1 mergers and blue points are
2:1 mergers. The magenta filled points show disky equal-mass merger remnants while magenta
open points correspond to boxy equal-mass mergers.}
\end{figure}

A detailed investigation of the intrinsic orbital structure of the merger remnants
and their global photometric and kinematical properties 
has been presented elsewhere (Naab \& Burkert 2001, 2003; Jesseit, Naab \& Burkert
2005).  Here we focus on their anisotropy. Figure 2 shows the location
of disky ($a_4 > 0.5$) equal- and unequal mass merger remnants in the anisotropy diagram. 
Like the observations, we here plot (v/$\sigma$)$_{1d}$ of the projected systems. The 
remnants have been rotated such that they are seen edge-on ($\vartheta=0$)
and are analysed with the line-of-sight parallel 
to their intermediate axis. Almost all unequal-mass mergers are disky. The situation is different for
equal-mass mergers. Only a few 1:1 mergers produced disky remnants (filled magenta points) for edge-on projections.
For comparison we also plot the boxy 1:1 mergers as open magenta points.

\begin{figure}
\vspace{1cm}
\centerline{\includegraphics[angle=0,scale=0.37]{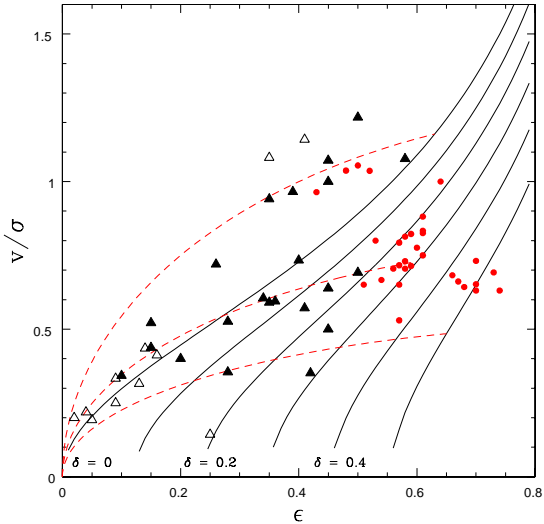}}
\caption{3:1 merger remnants of
disk galaxies are shown. The progenitors, in addition to the stellar component
also contained a gaseous disk with a gas-to-star mass fraction of 10\%. }
\end{figure}

The disky unequal-mass mergers are obviously not isotropic, in contrast to previous claims.
They fill the whole region between $\delta = 0$ and $\delta = 0.5$. In fact, our sample contains 
a larger fraction of anisotropic 4:1 mergers with $\delta \geq 0.4$ than is found for 1:1 mergers. 
Note also that there are almost no 3:1 to 4:1 mergers with $\delta = 0.3-0.4$. This might be due to 
the still limited coverage of the parameter space of initial disk inclinations.
The precise location of a merger remnant in the anisotropy diagram appears to
depend critically on the initial disk orientations. We will investigate this 
interesting sensitivity on the initial conditions in greater details in a subsequent paper. 
In summary, the main difference between
equal and unequal mass merger remnants is not their anisotropy but rather their rotation. Given 
the anisotropy $\delta$, increasing the mass ratio $\eta$ will
increase the value of $(v/\sigma)$. In addition, for a given mass ratio, the more
anisotropic the system the smaller its $(v/\sigma)$.

Clearly, the observations of disky ellipticals are not in agreement with edge-on merger remnants.
One might argue that an additional dissipative gas component could 
affect the rotation and ellipticity.
In order to investigate this question, we calculated all 3:1 mergers again, including now an
additional gaseous disk in the progenitors with a mass fraction of $10 \%$ the stellar mass.
A detailed analyses of these and additional simulations with different gas fractions
will be presented in a subsequent paper. Here we 
just note that the gas has an important effect on the intrinsic orbital structure of the 
remnants as it prohibits the formation of box orbits, by this also
affecting the shape of the observable line-of-sight velocity distribution.
However, somewhat surprisingly, the distribution of the merger remnants in the
anisotropy diagram is not affected as is shown in Fig. 3.

\section{Inclination effects}

\begin{figure}
\vspace{1cm}
\centerline{\includegraphics[angle=0,scale=0.37]{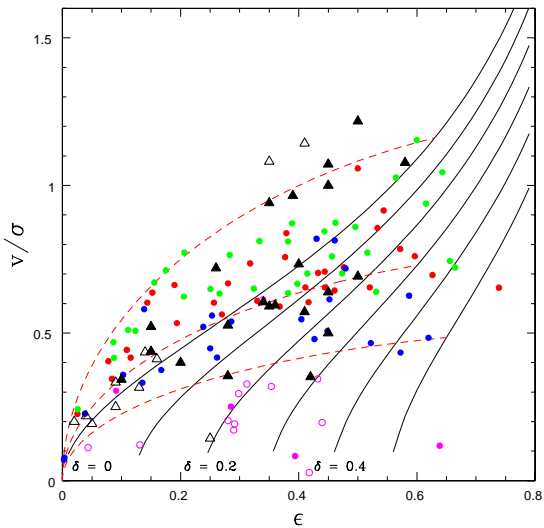}}
\caption{The distribution of projected merger remnants is compared with observed
disky ellipticals. The symbols are explained in Fig. 2.}
\end{figure}

As mentioned earlier, inclination effects can efficiently decrease the
apparent anisotropy of fast rotating ellipticals. 
The dashed lines in Fig. 2 show that inclining the merger remnants would indeed move them 
into the region occupied by the observations. To investigate this effect
more quantitatively, Fig. 4 shows again
the distribution of the merger remnants in the anisotropy diagram, however now inclined
with respect to the line-of-sight. For each object an inclination angle $\vartheta$
was randomly chosen with a probability $p(\vartheta) \ d \vartheta = \sin \vartheta \ d \vartheta$.
With a few exceptions, the distribution of merger remnants is in very
good agreement with the observations.  It might be interesting to note 
that the observed ellipticals appear to cluster in two groups. Group 1 is fast
rotating with $(v/\sigma) \approx 1$. This group can be barely fitted even by 4:1 merger remnants
(Naab \& Burkert 2003, Cretton et al. 2001).
Group 2 is characterized by $0.4 \leq (v/\sigma) \leq 0.7$ and can be well explained by inclined
2:1 - 4:1 mergers.  1:1 mergers are rotating too slowly to fit 
disky ellipticals, independent of inclination.

\begin{figure}
\vspace{1cm}
\centerline{\includegraphics[angle=0,scale=0.37]{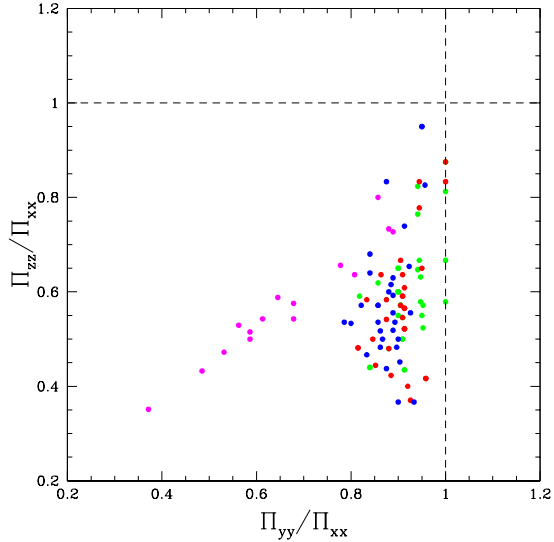}}
\caption{The ratios of the diagonal elements of the velocity
dispersion tensor are shown for equal- and unequal-mass merger remnants.
The symbols are described in Fig. 2.}
\end{figure}

\section{The intrinsic anisotropy of fast rotating merger remnants}

Is the anisotropy $\delta$ as derived from equation 21, using the observables
$v_{maj}$, $\sigma_0$ and $\epsilon$
a good estimate of the intrinsic anisotropy of fast rotating
stellar systems? To investigate this question, we have calculated the diagonal components
$\Pi_{ii}$ of the intrinsic velocity dispersion tensor (equation 6) of our merger remnants.
Figure 5 shows the relation between the ratios of the diagonal elements for all mergers.
In agreement with our previous assumption 
$\Pi_{xx} \approx \Pi_{yy} > \Pi_{zz}$ for unequal mass remnants. In contrast, boxy 
equal-mass mergers are characterized by $\Pi_{xx} >  \Pi_{yy} \approx \Pi_{zz}$.

Let us define the intrinsic anisotropy of an oblate merger remnant as
$1 - 2 \Pi_{zz}/(\Pi_{xx}+\Pi_{yy})$. The upper panel of Fig. 6 compares this quantity with
the "observationally" determined anisotropy $\delta$, derived from $(v/\sigma)_{1d}$ (Eq. 21).
The dashed curve shows the correlation expected
if both variables would agree. The merger remnants
have on average larger intrinsic anisotropies than
inferred from the projected properties.
The deviation increases with decreasing $\delta$.
Those ellipticals that appear isotropic actually have an intrinsic
anisotropy of 0.2, while ellipticals with apparently negative values of $\delta \approx -0.3$
are in reality isotropic.

\begin{figure}
\vspace{1cm}
\centerline{\includegraphics[angle=0,scale=0.65]{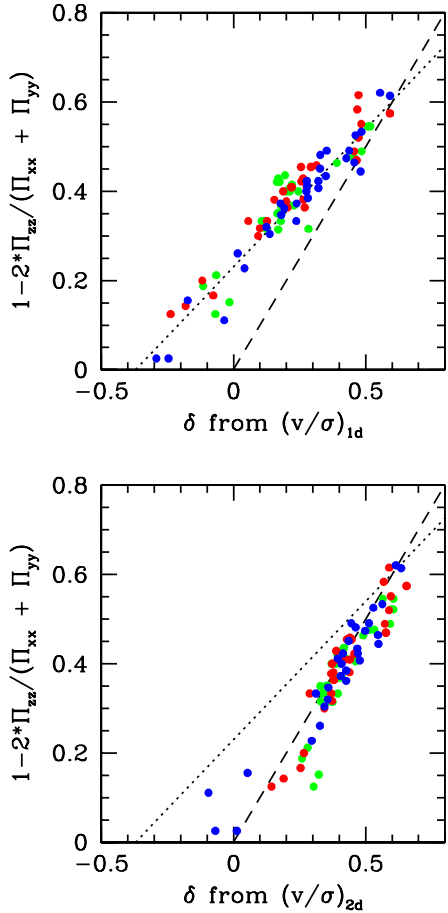}}
\caption{The upper panel compares the anisotropy $\delta$ of edge-on unequal-mass merger remnants, calculated
from the projected central velocity dispersion, characteristic major axis rotation 
and characteristic ellipticity (equ. 21) with their intrinsic anisotropy. The dotted line
shows a fit through the data (eq. 22). 
The lower panel compares the intrinsic anisotropy with $\delta$,
derived from equation 14 using the mean-squared ordered motion and velocity dispersion
of the projected, edge-on system and adopting $\alpha = 0$. The symbols are described in Fig. 2.}
\end{figure}

The knowledge of just three characteristic parameters ($v_{maj}$, $\sigma_0$, $\epsilon$) 
therefore seems not to be
sufficient to determine the intrinsic anisotropy of disky ellipticals accurately. 
One possible improvement is the spectroscopic measurement of the
two-dimensional projected average velocity field $\bar{v}_{los}$(x,z) and the
two-dimensional projected distribution of the velocity dispersion
$\sigma^2_{los}$(x,z). From this the mean-squared ordered motion 
$\langle \bar{v}^2_{los} \rangle$ (eq. 11) and the mean squared velocity
dispersion $\langle \sigma^2_{los} \rangle$ (eq. 12) could be derived and, given $\epsilon$,
equation (14) leads to a more precise determination of $\delta$. A crucial and uncertain
parameter is the value of $\alpha$ which  Binney (2005) estimates to be of order 0.13.
We have calculated the $\alpha$ of our merger remnants and find values 
of order 0.08 to 0.2 for 3:1 and 4:1 remnants with somewhat larger values for
2:1 mergers. $\alpha$ tends to increase with decreasing $(v/\sigma)$ and 
decreasing progenitor mass ratio $\eta$. However the total number of particles
used in our simulations is too small to determine $\alpha$  accurately and simulations
with substantially larger particle numbers would be required. Interestingly, using $\alpha = 0$
already leads to a reasonably good estimate of the intrinsic anisotropy. To demonstrate this,
the lower panel of Fig. 6 plots $\delta$, derived from $(v/\sigma)_{2d}$ with $\alpha = 0$,
using equation 14 and compares it with the intrinsic anisotropy of the merger remnants.
The agreement is much better than previously.

The upper panel of Fig. 6 shows that all 
fast rotating, disky ellipticals follow
a narrow correlation between the intrinsic and observationally inferred 
anisotropy which can be well fitted by the empirical formula (dotted line)

\begin{equation}
1 - 2 \Pi_{zz}/(\Pi_{xx}+\Pi_{yy}) = 0.62 \ \delta + 0.23   .
\end{equation}

This suggests a second, less expensive method to determine the intrinsic anisotropy from
the usually measured observables: $\epsilon$, $\sigma_0$ and $v_{maj}$.
Using equation 21, a first estimate of $\delta$ is derived.
Equation 22 then gives the intrinsic anisotropy. Note however that all of these
considerations require first a correction due to inclination. 
Unless the inclination angle $\vartheta$ of the system
is known, e.g. by measuring the orientation of an additional central disk component,
inclination effects will dominate the uncertainties and will make a precise
determination of the intrinsic anisotropy impossible.

\section{Conclusions}

Numerical simulations of unequal-mass disk galaxy mergers lead to fast rotating 
stellar systems that resemble observed disky elliptical galaxies. However, in 
contrast to previous claims, the objects have a large spread in their
anisotropies,
ranging from almost isotropic systems to objects with 
$1 - \Pi_{zz}/(\Pi_{xx}+\Pi_{yy}) \approx 0.65$. The distribution of anisotropies is
similar to boxy, slowly rotating merger remnants that result from equal-mass mergers 
and that are believed to explain the origin of massive boxy ellipticals.
The main difference between disky and boxy ellipticals therefore is not anisotropy
but the amount of rotation and the fact that unequal-mass mergers are characterised by
$\Pi_{xx} \approx \Pi_{yy} > \Pi_{zz}$ whereas 
equal-mass mergers have $\Pi_{xx} > \Pi_{yy} \approx \Pi_{zz}$.

Figure 2 shows that the ellipticity of anisotropic fast rotating merger remnants 
is larger than expected for the isotropic case which means that the 
flattening of these objects is at least partly a result of their anisotropic 
velocity distribution. Including gas does not change this conclusion. 
For example, 3:1 and 4:1 merger
remnants with $\delta \approx 0.5$ and $v/\sigma \approx 0.7$ have ellipticities 
of $\epsilon \approx 0.7$, whereas isotropic objects with similar
values of $v/\sigma$ would be much rounder with $\epsilon \approx 0.4$.
Inclination effects strongly reduce the apparent anisotropy, especially for fast 
rotating systems. For example, Fig. 1 shows that objects with $v/\sigma \approx 1$
and $\delta = 0.5$ would already appear isotropic with $\delta = 0$ when viewed under an 
inclination angle of $\vartheta = 30^{\circ}$. This explains why fast
rotating disky ellipticals appear isotropic and reconciles the predictions of
the major merger scenario with the observations. Clearly,
more work on determining inclination angles for fast rotating ellipticals is
required to gain a deeper insight into their internal dynamical structure
and their origin.
\newline

We thank James Binney for sharing his new analytical calculations on
the anisotropy determination of projected stellar systems with us
prior to publication. We also thank Ralf Bender for
many interesting discussions.



\label{lastpage}

\end{document}